\documentclass[traditabstract]{aa} 
\usepackage{graphicx}
\usepackage{natbib}
\usepackage{color}

\bibpunct{(}{)}{;}{a}{}{,} 
\usepackage{mathcomp}
\usepackage{graphicx}
\usepackage{txfonts}
\usepackage{longtable}
\usepackage{lscape}
\usepackage{wasysym}

\newcommand{\Mearth}{$\mathrm{M_\oplus}$}

\newcommand{\AU}{{\sc au}}
\newcommand{\etal}{et al.}

\newcommand{\hst}{\emph{HST}}
\newcommand{\chandra}{\emph{Chandra}}
\newcommand{\xmm}{\emph{XMM-Newton}}
\newcommand{\lya}{Ly$\alpha$}
\newcommand{\kms}{km~s$^{-1}$}
\begin{document}

	\title{Hint of a transiting extended atmosphere on 55~Cancri~b\thanks{Based on observations made with the NASA/ESA \emph{Hubble Space Telescope}, obtained at the Space Telescope Science Institute, which is operated by the Association of Universities for Research in Astronomy, Inc., under NASA contract NAS 5-26555. These observations are associated with GO/DD programme \#12681.}}

   \author{D.~Ehrenreich\inst{1,2}, V.~Bourrier\inst{3}, X.~Bonfils\inst{2}, A.~Lecavelier des Etangs\inst{3}, G.~H\'ebrard\inst{3,4}, D.~K.~Sing\inst{5}, P.~J.~Wheatley\inst{6}, A.~Vidal-Madjar\inst{3}, X.~Delfosse\inst{2}, S.~Udry\inst{1}, T.~Forveille\inst{2} \& C.~Moutou\inst{7}}
          
   \institute{
	Observatoire astronomique de l'Universit\'e de Gen\`eve, Sauverny, Switzerland\\     	\email{david.ehrenreich@unige.ch}
	\and
    UJF-Grenoble 1 / CNRS-INSU, Institut de Plan\'etologie et d'Astrophysique de Grenoble (IPAG) UMR 5274, Grenoble, France
    \and
	Institut d'astrophysique de Paris, Universit\'e Pierre \& Marie Curie, CNRS (UMR~7095), Paris, France
	\and
	Observatoire de Haute-Provence, CNRS (USR~2217), Saint-Michel-l'Observatoire, France
	\and
	Astrophysics Group, School of Physics, University of Exeter, Exeter, UK
	\and
	Department of Physics, University of Warwick, Coventry, UK
	\and
	Laboratoire d'astrophysique de Marseille, Universit\'e de Provence, CNRS (UMR~6110), Marseille, France
	}

   \date{}
 
  \abstract{
The naked-eye star \object{55~Cancri} hosts a planetary system with five known planets, including a hot super-Earth (\object{55~Cnc~e}) extremely close to its star and a farther out giant planet (\object{55~Cnc~b}), found in milder irradiation conditions with respect to other known hot Jupiters. This system raises important questions on the evolution of atmospheres for close-in exoplanets, and the dependence with planetary mass and irradiation. These questions can be addressed by Lyman-$\alpha$ transit observations of the extended hydrogen planetary atmospheres, complemented by contemporaneous measurements of the stellar X-ray flux. In fact, planet `e' has been detected in transit, suggesting the system is seen nearly edge-on. Yet, planet `b' has not been observed in transit so far. Here, we report on \emph{Hubble Space Telescope} STIS \lya\ and \chandra\ ACIS-S X-ray observations of 55~Cnc. These simultaneous observations cover two transits of 55~Cnc~e and two inferior conjunctions of 55~Cnc~b. They reveal the star as a bright \lya\ target and a variable X-ray source. While no significant signal is detected during the transits of 55~Cnc~e, we detect a surprising \lya\ absorption of $7.5\pm1.8\%$ (4.2~$\sigma$) at inferior conjunctions of 55~Cnc~b. The absorption is only detected over the range of Doppler velocities where the stellar radiation repels hydrogen atoms towards the observer. We calculate a false-alarm probability of 4.4\%, which takes into account the a-priori unknown transit parameters. This result suggests the possibility that 55~Cnc~b has an extended upper \ion{H}{i} atmosphere, which undergoes partial transits when the planet grazes the stellar disc. If confirmed, it would show that planets cooler than hot Jupiters can also have extended atmospheres.}


   \titlerunning{Hint of a transiting extended atmosphere on 55~Cnc~b}
   \authorrunning{Ehrenreich \etal}

   \maketitle
%

\section{Introduction}

A large fraction ($\sim$27\%) of exoplanets are found at close separations ($0.01$--$0.1$~\AU) from their host stars, as a result of a detection bias.\footnote{The velocimetric and photometric detection techniques are  more efficient for short-period planets.} These hot planets sustain high levels of extreme ultraviolet (EUV) and X-ray radiations, which can impact their atmospheric properties, dynamics, and ultimately their fate. Observations of transits in the stellar Lyman-$\alpha$ (\lya) emission line of neutral hydrogen (\ion{H}{i}) have revealed extended upper atmospheres on the hot giant exoplanets \object{HD~209458b} \citep{Vidal-Madjar:2003,Vidal-Madjar:2004b,Vidal-Madjar:2008,Ben-Jaffel:2007,Ben-Jaffel:2008,Ehrenreich:2008} and \object{HD~189733b} \citep{LecavelierdesEtangs:2010,LecavelierdesEtangs:2012}. These extended \ion{H}{i} envelopes are also observed along with heavier atoms and ions \citep{Vidal-Madjar:2004b,Ben-Jaffel:2010,Linsky:2010}, also possibly detected on \object{Wasp-12b} \citep{Fossati:2010}. The escape of heavy elements suggests the upper atmospheres are in a hydrodynamical `blowoff'\footnote{A discussion about the possible meanings of `blowoff' can be found in \citet{Tian:2008}} state \citep{GarciaMunoz:2007,Murray-Clay:2009}. Such hydrodynamic escape is thought to have occurred for much cooler planets in the early Solar System \citep{Lammer:2008}. This raises two interesting questions: (i) Is there a critical distance to the star below which an atmosphere would experience a sustained hydrodynamical escape \citep{Koskinen:2007b}? (ii) Is there a critical mass threshold below which a planet can be totally stripped of its atmosphere \citep{LecavelierdesEtangs:2004,Owen:2012}?

The 55~Cancri planetary system can help address both questions. This star (G8V) is nearby (12.3~pc), bright ($V=5.95$), and a host to five exoplanets \citep{Fischer:2008}. Planet `e' is the closest to the star, a hot super-Earth ($M_e=7.81^{+0.58}_{-0.53}$~\Mearth) at only 0.0156~\AU\ \citep{Dawson:2010}, recently detected in transit \citep{Winn:2011, Demory:2012}. Planet `b' is the second closest, a Jupiter-mass object ($M_b = 0.8$~M$_\textrm{\jupiter}$) at 0.115~\AU, close to the predicted limit of 0.15~\AU\  where `a sharp breakdown in atmospheric stability' should occur for jovian planets around solar-type stars \citep{Koskinen:2007b}. Studying the exospheric properties of 55~Cnc~e and b could therefore provide key insights into the critical mass and distance at which atmospheric escape may dramatically impact the atmospheric evolution of exoplanets. However, 55~Cnc~b has not been detected in transit so far.

In this article, we report on \emph{Hubble Space Telescope} (\hst) \lya\ observations of transits of 55~Cnc~e and inferior conjunctions of 55~Cnc~b (Sect.~\ref{sec:obs}). They are complemented by contemporaneous \chandra\ observations aimed at monitoring the X-ray irradiation received by the planets. While the \lya\ observations show no evidence for a transit signal during the transits of planet~e, they hint at a transit-like signal detected in phase with inferior conjunctions of planet~b (Sect.~\ref{sec:results}), suggesting that the extended upper atmosphere of 55~Cnc~b undergoes partial transits (Sect.~\ref{sec:discussion}).

\section{Observations and data analysis}
\label{sec:obs}

\subsection{\emph{Hubble Space Telescope} \lya\ observations}
We obtained \hst\ time (GO/DD~12681) to observe 55~Cnc during two visits on 2012-03-07 and 2012-04-05 with the Space Telescope Imaging Spectrograph \citep[STIS;][]{Woodgate:1998}. Each visit consists of four consecutive \hst\ orbits. The data were recorded in time-tag mode with exposure times of 2\,040 to 3\,005~s on the Far Ultraviolet Multi-Anode Microchannel Array (FUV-MAMA) detector. Incoming light was diffracted with the long $52\arcsec \times 0\farcs1$ slit and the G140M grating, yielding first-order spectra centered on 1\,222~\AA\ and ranging from 1\,194 to 1\,249~\AA. The medium spectral resolution obtained is $\sim10\,000$ ($\sim30$~\kms).

\begin{figure}
\resizebox{\columnwidth}{!}{\includegraphics[trim=2cm 0.5cm 1cm 0.5cm]{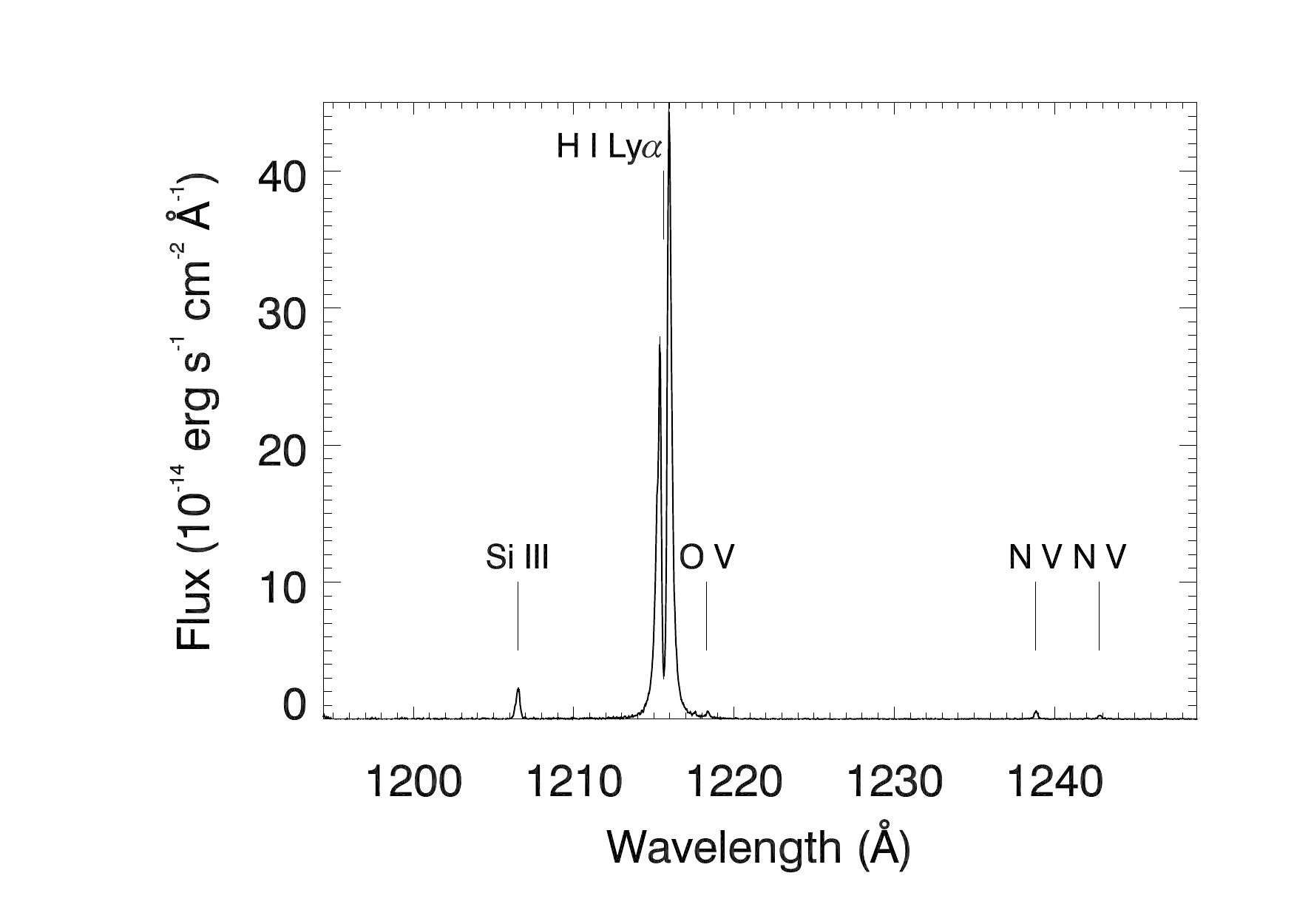}}
\caption{\label{fig:FullSpectrum} STIS/G140M spectrum of 55~Cnc.}
\end{figure}

\begin{figure}[!t]
\resizebox{\columnwidth}{!}{\includegraphics[trim=2cm 0.5cm 1cm 0.5cm]{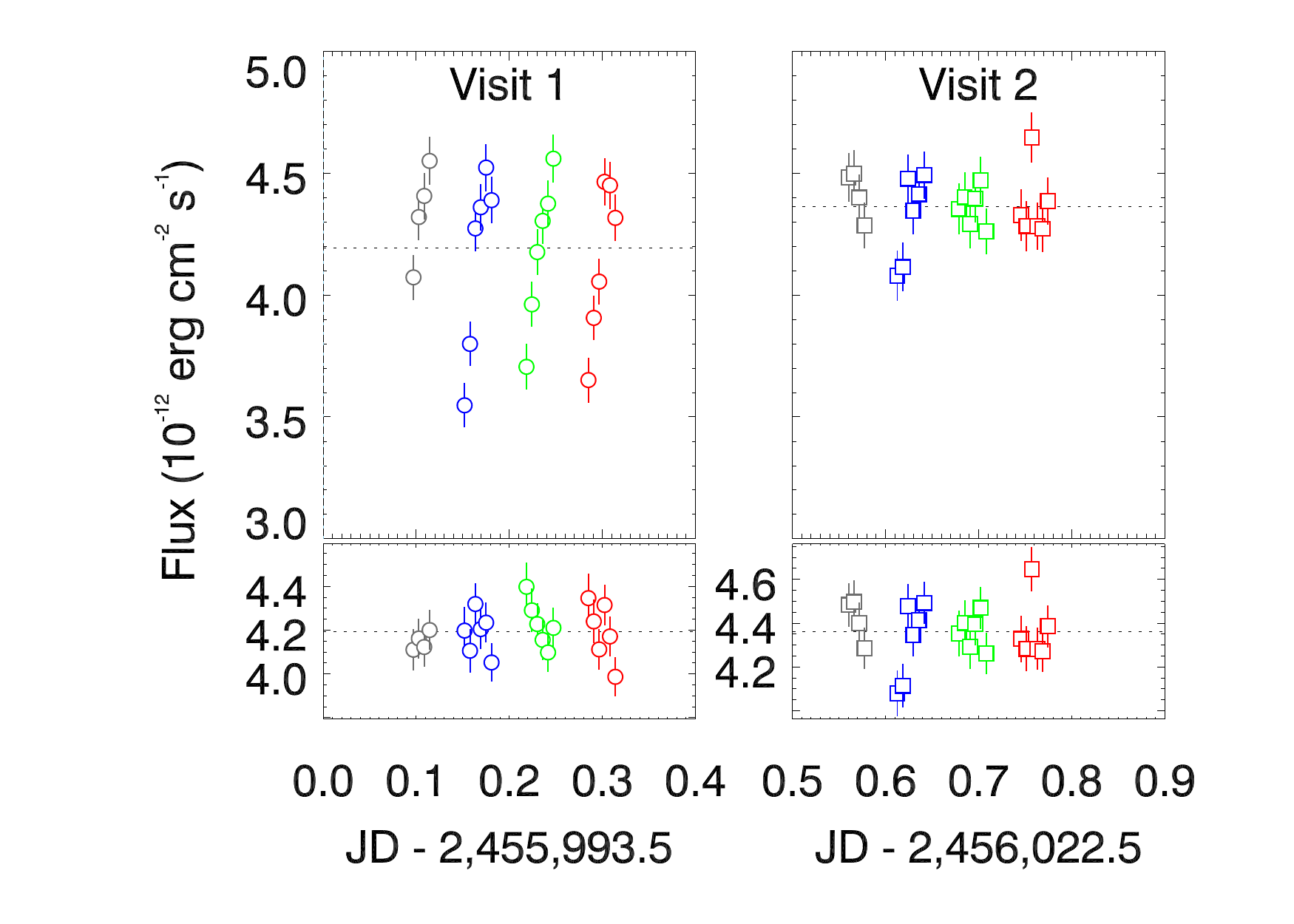}}
\caption{\label{fig:LCtot} \emph{Top panels:} Lyman-$\alpha$ flux integrated over the whole \lya\ line as a function of time for 500-s sub-exposures in visits~1 (circles, \emph{left}) and~2 (squares, \emph{right}). The different colours mark the different \hst\ orbits. \emph{Bottom panels:} Lyman-$\alpha$ flux corrected for the `breathing effect'. Dotted lines show the mean fluxes in visits~1 and~2.}
\end{figure}

\begin{figure}
\resizebox{\columnwidth}{!}{\includegraphics[trim=2cm 0.5cm 1cm 0.5cm]{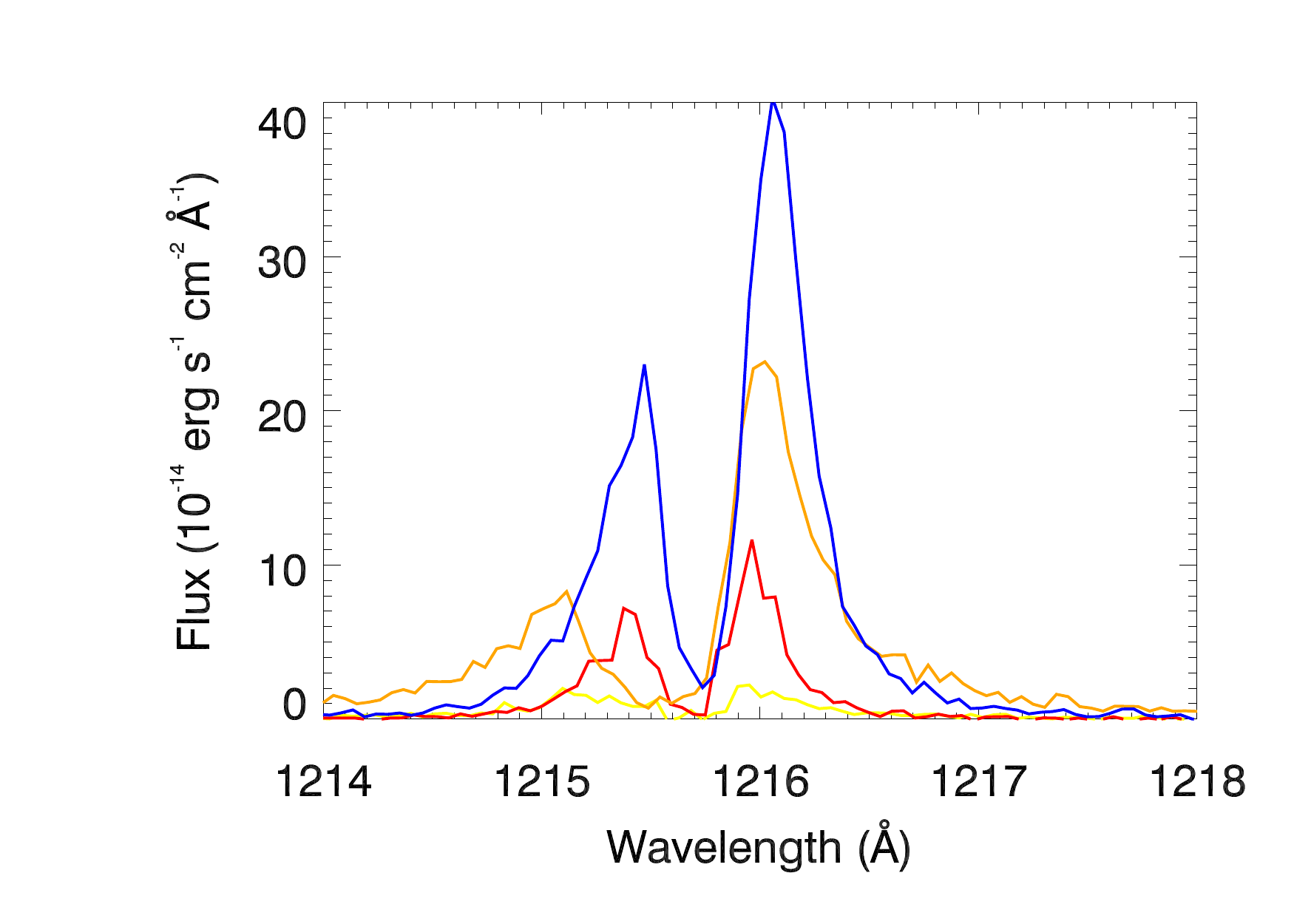}}
\caption{\label{fig:StellarLyaEmissions} Comparison of STIS/\lya\ spectra of stars hosting transiting exoplanets: the G8V star 55~Cnc at 12.3~pc (this work, blue), the K1--K2  star HD~189733 at 19.3~pc \citep[orange]{LecavelierdesEtangs:2012}, the M3.5V star GJ~436 at 10.2~pc \citep[red]{Ehrenreich:2011a}, and the G0V star HD~209458 at 49.6~pc \citep[yellow]{Vidal-Madjar:2003}.}
\end{figure}

Figure~\ref{fig:FullSpectrum} shows an example of a STIS/G140M spectrum of 55~Cnc. The \lya\ line (1\,215.67~\AA) is, by far, the most prominent feature in the whole emission spectrum. Between 1\,214.5 and 1\,217~\AA, we measured mean \lya\ fluxes of $(4.19\pm0.02)\times10^{-12}$ and $(4.36\pm0.02)\times10^{-12}$~erg~cm$^{-2}$~s$^{-1}$ in visits~1 and~2, respectively. The quoted uncertainties are provided by the STIS pipeline. They are not representative of the observed flux scatter, as shown in Fig.~\ref{fig:LCtot} and discussed in the next paragraph. These \lya\ flux levels are brighter than for any other known transit-hosting stars, as shown in Fig.~\ref{fig:StellarLyaEmissions}. The \lya\ feature has a typical double peak due to absorption in the line centre by the interstellar \ion{H}{i} along the line of sight. Here, the interstellar absorption is narrow, and in particular the `blue peak' of the line is much less absorbed by the interstellar medium than in the case of HD~189733, allowing us to probe the line at lower Doppler velocities. Other stellar emission from ionised silicon (\ion{Si}{iii} at 1\,206.5~\AA), oxygen (\ion{O}{v} at 1\,218.3~\AA), and nitrogen (\ion{N}{v} doublet at 1\,238.8 and 1\,242.8~\AA) are also detected. These lines do not show any significant variations during the observations and in the following, we will  dedicate our attention to the \lya\ line.

\begin{figure}[!t]
\resizebox{\columnwidth}{!}{\includegraphics[trim=2cm 3cm 1cm 0.5cm]{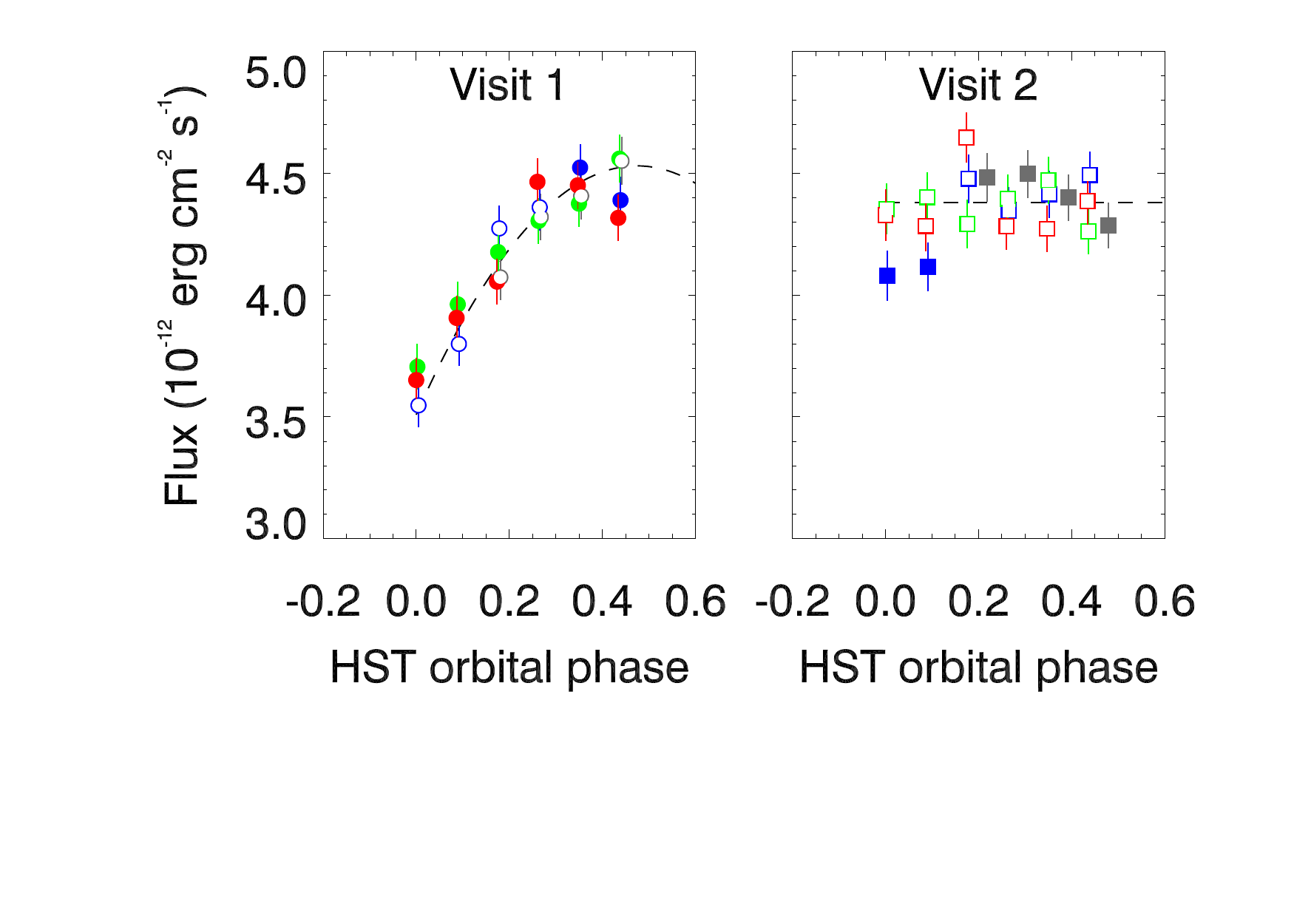}}
\caption{\label{fig:breathing} Lyman-$\alpha$ flux phase-folded on the \hst\ orbital period. Filled symbols are exposures taken near the inferior conjunction of 55~Cnc~b, which are excluded from the fit. The dashed lines indicate the best-fit quadratic function for visit~1 (`breathing effect'; \emph{left}) and constant for visit~2 \emph{(right)}.}
\end{figure}

Taking advantage of the time-tag data acquisition mode, each orbit-long exposure was split into 500-s sub-exposures, for a total of 22 sub-exposures per visit. This enabled a monitoring of the \lya\ flux during each \hst\ orbit (Fig.~\ref{fig:LCtot}), which revealed a systematic effect clearly seen in visit~1: the \lya\ flux increases with \hst\ orbital phase (Fig.~\ref{fig:breathing}), introducing a scatter of 7\% and 3\% around the average fluxes quoted above for visits~1 and~2, respectively. This effect (of a maximal amplitude of $\sim20\%$) is attributed to optical telescope assembly `breathing' -- secondary mirror motions due to orbital thermal variations \citep{Kimble:1998}. We determined that this effect can be modelled with a quadratic function of the orbital phase \citep[see also][]{Sing:2011b}. The fit was performed on a sample trimmed from the exposures obtained near the inferior conjunction of 55~Cnc~b, so as to avoid a possible signal contamination, yielding a $\chi^2$ of 3.85 for $\nu = 8-3$ degrees of freedom ($\chi^2_\nu=0.77$). 

In contrast, visit~2 is not impacted by telescope breathing. We checked that fitting linear or quadratic functions to the phase-folded \lya\ light curve (also trimmed from values close to the inferior conjunction of planet `b') does not statistically improve the dispersion measured in the `raw' light curve. Correction factors are derived as a function of the \hst\ orbital phase, as the ratio between the polynomial value to the mean polynomial value over the observed phase range. The detrended \lya\ light curves extracted from the corrected spectra are shown in Fig.~\ref{fig:LCtot}.

\subsection{\chandra\ X-ray observations}
\label{sec:chandra}
In order to monitor the stellar X-ray radiation, we obtained \chandra\ ACIS-S coverage \citep{Weisskopf:2002,Garmire:2003} of each \hst\ visit (\chandra\ ObsIDs 14401 \& 14402). The observations were scheduled to begin 4 hours before mid-transit of 55~Cnc~e, with precise start times of 2012-03-07 UT~02:13 and 2012-04-05 UT~12:50. The scheduled exposure times were 20~ks, but the first observation was cut short to 12~ks by safing of ACIS due to high Solar radiation levels. Since 55~Cnc is a bright optical source, we operated the ACIS-S CCD in very faint (VF) mode and selected a 1/8 sub-array in order to minimise optical loading.

We analysed the \chandra\ data using CIAO version 4.4. A weak soft X-ray source was detected at the proper-motion corrected co-ordinates of 55~Cnc in both observations. We extracted source counts using a region of radius 1\farcs7, and the background was estimated using two large circular regions of radius 27\arcsec\ adjacent to the target and aligned along the sub-array. The source had consistent background subtracted count rates in the two observations of $0.0019\pm0.0005$ and $0.0022\pm0.0004$~s$^{-1}$.  \chandra\ ACIS-S light curves of 55~Cnc are presented in Figs.~\ref{fig:LCb}~\&~\ref{fig:LCe}.

For comparison, we also analysed an archival \xmm\ X-ray observation of 55~Cnc from April 2009 (\xmm\ ObsID.\ 0551020801) previously reported by \citet{Sanz-Forcada:2011}. We used SAS version 12.0 and extracted source counts from the EPIC-pn camera using a circular region of radius 19\farcs2. The background was estimated from a large adjacent region on the same CCD selected to be free from other sources. The count rate was found to be $0.0187\pm0.0015\rm\,s^{-1}$.

\section{Results}
\label{sec:results}

\subsection{55~Cancri~b}

The orbital properties of 55~Cnc~b \citep{Butler:1997} were accurately constrained by \citet{Fischer:2008}, from 18 years of velocimetric measurements (from 1989 to 2007) obtained at the Lick and Keck observatories. These authors ruled out \emph{central} transits of planets b, c, and d. \citet{Dawson:2010} re-analysed the data and revised the orbital period of 55~Cnc~e, from 2.8 to 0.74 days. Transits of planet e have been subsequently detected \citep{Winn:2011,Demory:2012}, suggesting the whole system is seen nearly edge-on. There is thus a possibility that the second closest planet, 55~Cnc~b, could graze the stellar disc at its closest projected approach --\,the inferior conjunction.

\subsubsection{Constraints on the inferior conjunction times}

\citet{Dawson:2010} provide an ephemeris for 55~Cnc~b, predicting an inferior conjunction at BJD~$2\,453\,092.752\pm0.023$ (2004-03-28). To update their prediction as accurately as possible, we pooled together Keck and Lick \citep{Fischer:2008}, HET \citep{McArthur:2004}, and SOPHIE radial velocities (Moutou et al., in preparation). We performed a 5-planet Keplerian fit to these data and introduced a systematic `jitter' for each data set, which is free to vary and quadratically added to the quoted uncertainties on the radial velocities, so to weigh the relative contribution of each time series according to its true precision. We obtained for 55~Cnc~b an orbital period of $14.651\,364\pm0.000\,074$~days, and predict inferior conjunctions during \hst\ visits on BJD~$2\,455\,993.770\pm0.013$ (2012-03-07 UT 06:28$\pm19$~min) and BJD~$2\,456\,023.073\pm0.013$ (2012-04-05 UT 13:44$\pm$19~min).

\begin{figure}
\resizebox{\columnwidth}{!}{\includegraphics[trim=2.cm 0.5cm 1cm 0.5cm]{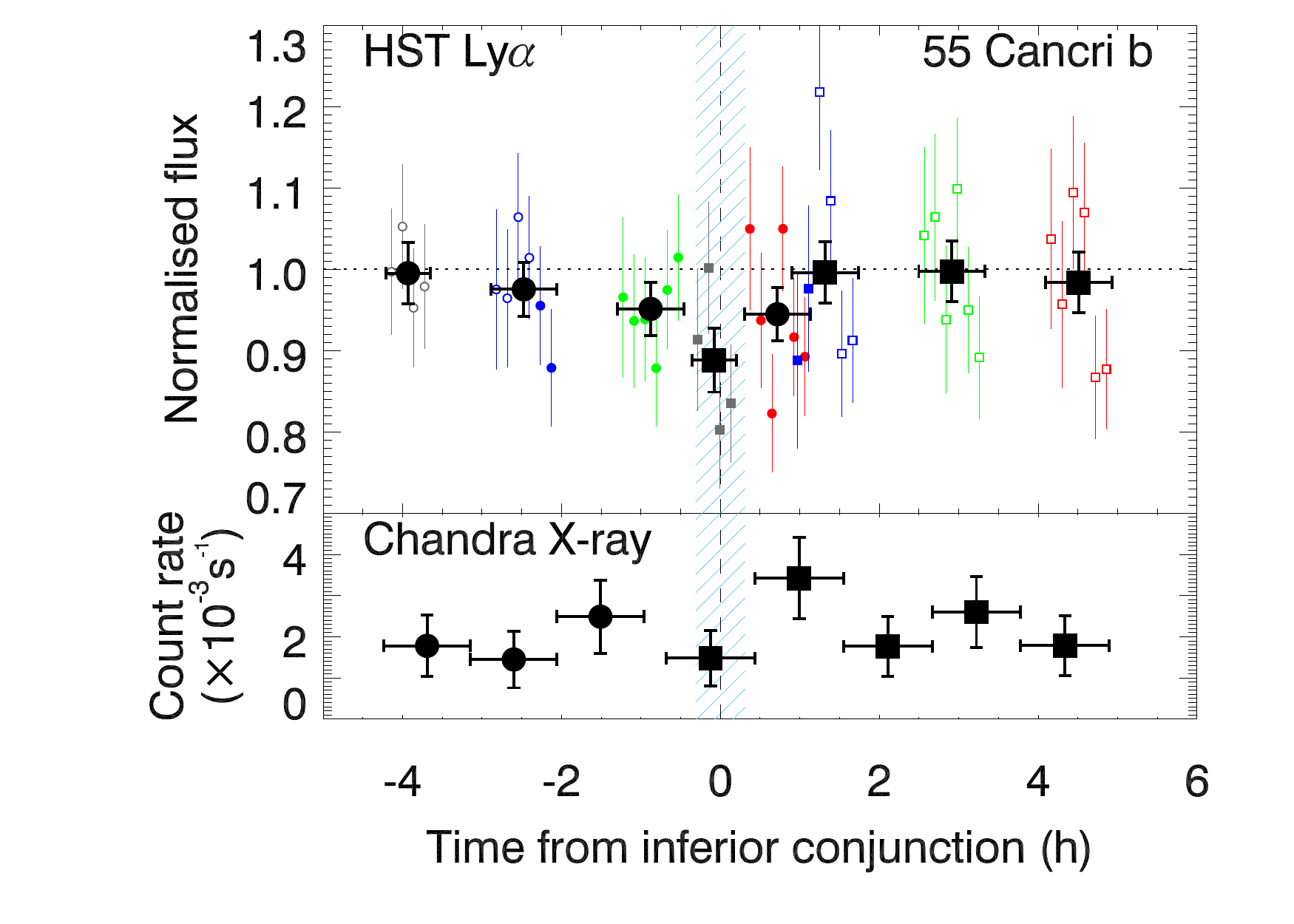}}
\caption{\label{fig:LCb} \emph{Top.} Lyman-$\alpha$ light curve integrated between $1\,215.36$ and $1\,215.67$~\AA\ or between $-76.5$ and $0$~\kms. The exposures from visits~1 (small circles) and~2 (small squares) have been phase-folded around the period of 55~Cnc~b. The different colours (grey, blue, green, red) correspond to the four orbits in each visit. Exposures considered in and out of transit are differenciated by filled and empty symbols, respectively. The large black symbols (circles for visit~1, squares for visit~2) are the flux averages per orbit. The sky-blue hatched region shows the $\pm19$-min uncertainty on the inferior conjunction time. \emph{Bottom.} \chandra\ X-ray count rate integrated over $\sim1$-h bins.}
\end{figure}

Combining both \hst\ visits provides an excellent coverage of the inferior conjunction of 55~Cnc~b. As for HD~209458b and HD~189733b, we would expect any transit signatures as absorption over a limited and blueshifted range of velocities in the \lya\ line. We scanned the spectra in search for transit-like signatures in the vicinity of the inferior conjunction. For each wavelength range, we calculated a phase-folded transit light curve. Since the average \lya\ flux level varies between the two epochs (see Fig.~\ref{fig:LCtot}), the fluxes are normalised in each visit by the mean `out-of-transit' flux (which is determined a posteriori). At the predicted time of the inferior conjunction, we found a blueshifted absorption signal of $7.5\pm1.8\%$ between $1\,215.36$ and $1\,215.67$~\AA, or between Doppler velocities of $-76.5$ and $0$~\kms, as can be seen on the phase-folded light curve (Fig.~\ref{fig:LCb}) and on the transmission spectrum (Fig~\ref{fig:Spectra_b}).

\subsubsection{Impact of background subtraction}

This wavelength range is contaminated by the terrestrial air glow, which shows large --\,yet reproducible\,-- variations during the \hst\ orbits within a visit. The air glow impact on \lya\ transits has been extensively discussed in the case of HD~209458 \citep{Vidal-Madjar:2008,Ben-Jaffel:2007}. However, unlike HD~209458 where the \lya\ line is about as bright as the air glow, the \lya\ emission of 55~Cnc is $\sim20$ times brighter than the air glow. We nevertheless performed a battery of tests to assess the possible impact of the air glow subtraction to the reduced spectra. The STIS reduction pipeline takes advantage of the spatial extent of the air glow on the detector $y$-axis to subtract it, together with the background. The background level is estimated in a region located $\pm300$ pixels away from the centre of the spectral extraction region \citep{Bostroem:2011}. \citet{Ben-Jaffel:2007} called attention to the possible flux variations on the detector along the $y$-axis, which represent --\,according to this author\,-- an incompressible uncertainty of 5\% on the flux level of extended sources. We mapped the air glow along the $y$-axis for each \hst\ orbit and found that systematic variations are indeed present. To quantify their effect, we ran several data reductions, extracting the background at different locations along the detector $y$ axis. These tests show that the air glow --\,and more generally the background\,-- subtraction (i) does not impact the detection of a transit signal for 55~Cnc~b, yet (ii) induces a mean dispersion of $\sim3\%$ on the flux measured on each 500-s sub-exposure. This dispersion was quadratically added to the flux uncertainties, and propagated throughout our calculations. 

\subsubsection{Transit-like signal and false-alarm probability}

In order to estimate the mean `in-transit' absorption, we need to define the transit contacts, i.e., which exposures are `in transit' and which ones are `out of transit'. Because the transit is not detected in the optical and the \lya\ transit duration is not known, a systematic exploration of possible start and end times of the transit has been undertaken. The largest absorption signal ($7.5\pm1.8\%$) is obtained from $\sim2.3$~h before to $\sim1$~h after the inferior conjunction of 55~Cnc~b.

False-alarm probabilities (FAPs) should be associated with the measured signal, to take the choices of a-priori unknown parameters into account. These choices are (i) the transit start (${T_\textrm{\sc i}}_b$) and end (${T_\textrm{\sc iv}}_b$) times, and (ii) the wavelength domain in the \lya\ line. The FAP associated with (i) is calculated with a boot-strap method, randomising the choices of ${T_\textrm{\sc i}}_b$ and ${T_\textrm{\sc iv}}_b$. In each trial, the transit depth is estimated and the associated signal-to-noise ratio is compared to the fiducial value ($S/N=4.2$). After $10^5$ trials, we found a FAP of 1.8\%. Next, to estimate the FAP associated with (ii), we generated new spectra for all of the 500-s exposures, combining a reference averaged spectrum with random noise. Then, we searched the blue part of the \lya\ line in generated spectra for a wavelength region where the in-transit absorption (that would be due here to noise flucutation) would be more significant than the fiducial signal. For $5\times10^4$ trials, we found a FAP of 2.6\%. Since these FAPs are independent, the final probability that the signal is real is $(1-0.018)\times(1-0.026)=95.6\%$ (FAP of 4.4\%).

\begin{figure*}
\resizebox{\textwidth}{!}{\includegraphics{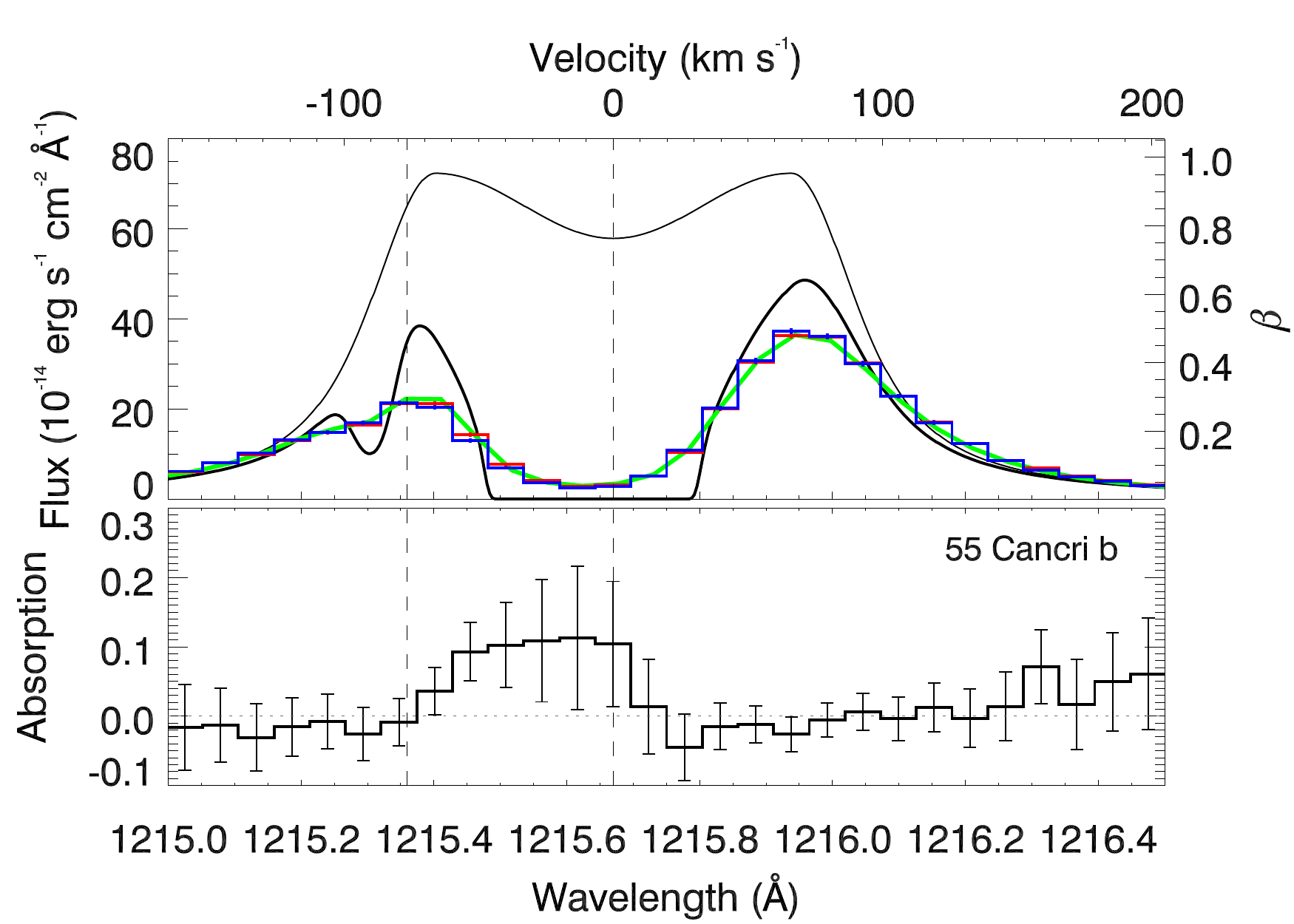}}
\caption{\label{fig:Spectra_b} \emph{Top.} Lyman-$\alpha$ line of 55~Cnc. Weighted co-addition of exposures flagged `in' ($F_\mathrm{in}$; solid blue line) and `out' ($F_\mathrm{out}$; solid red line) of the possible transit of 55~Cnc~b's extended atmosphere. The solid black lines show the stellar intrinsic emission profile before (thin) and after (thick) absorption by the interstellar medium, in terms of flux (left axis) or $\beta$ ratio between radiation pressure and gravity (right axis). The absorbed profile is convoluted with the instrumental line spread function to yield the green profile, which is fit to the data. Dashed vertical lines delimit the region over which the flux is integrated to produce the light curve in Fig.~\ref{fig:LCb}.
\emph{Bottom.} Transmission spectrum showing the absorption depth $\delta = 1 - F_\mathrm{in}/F_\mathrm{out}$ during the possible transit of 55~Cnc~b.}
\end{figure*} 

\subsection{55~Cancri~e}

\begin{figure}
\resizebox{\columnwidth}{!}{\includegraphics[trim=2.cm 0.5cm 1cm 0.5cm]{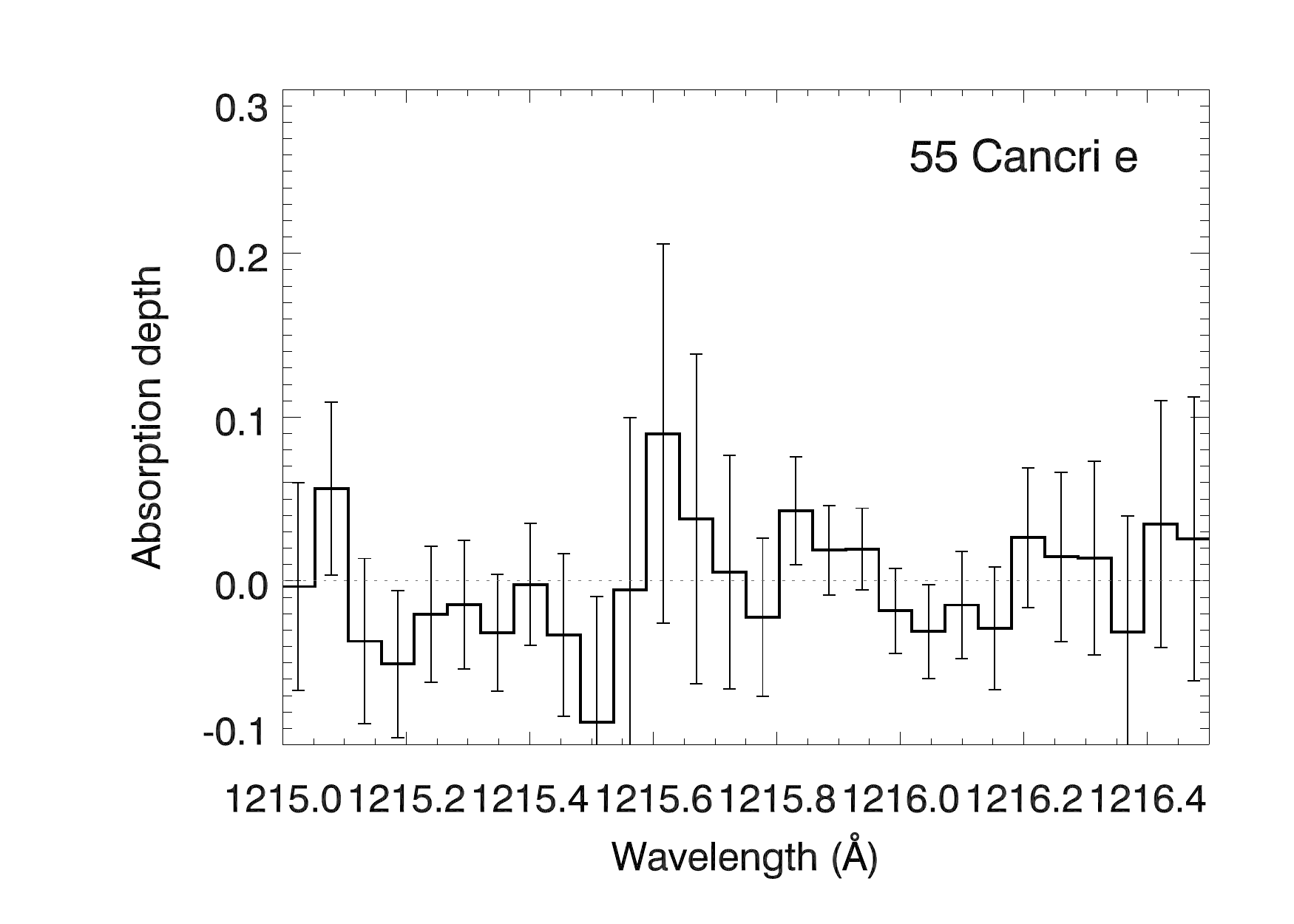}}
\caption{\label{fig:AbsorptionDepth} Absorption depths during transits of 
55~Cnc~e, as a function of wavelength along the \lya\ line.}
\end{figure}

A similar analysis is performed for 55~Cnc~e. Figure~\ref{fig:AbsorptionDepth} shows the absorption depth $\delta = 1 - F_\mathrm{in}/F_\mathrm{out}$ as a function of wavelength (the transmission spectrum) for planet~e. The `best' signal, $7.0\pm5.3$\%, is found between 1\,215.56 and 1\,215.67~\AA\ or between $-26.6$ and $0$~\kms, with an associated FAP of 89\%. The weak significance of this signal coupled with the high FAP call for a spurious origin.

\begin{figure}
\resizebox{\columnwidth}{!}{\includegraphics[trim=2.cm 0.5cm 1cm 0.5cm]{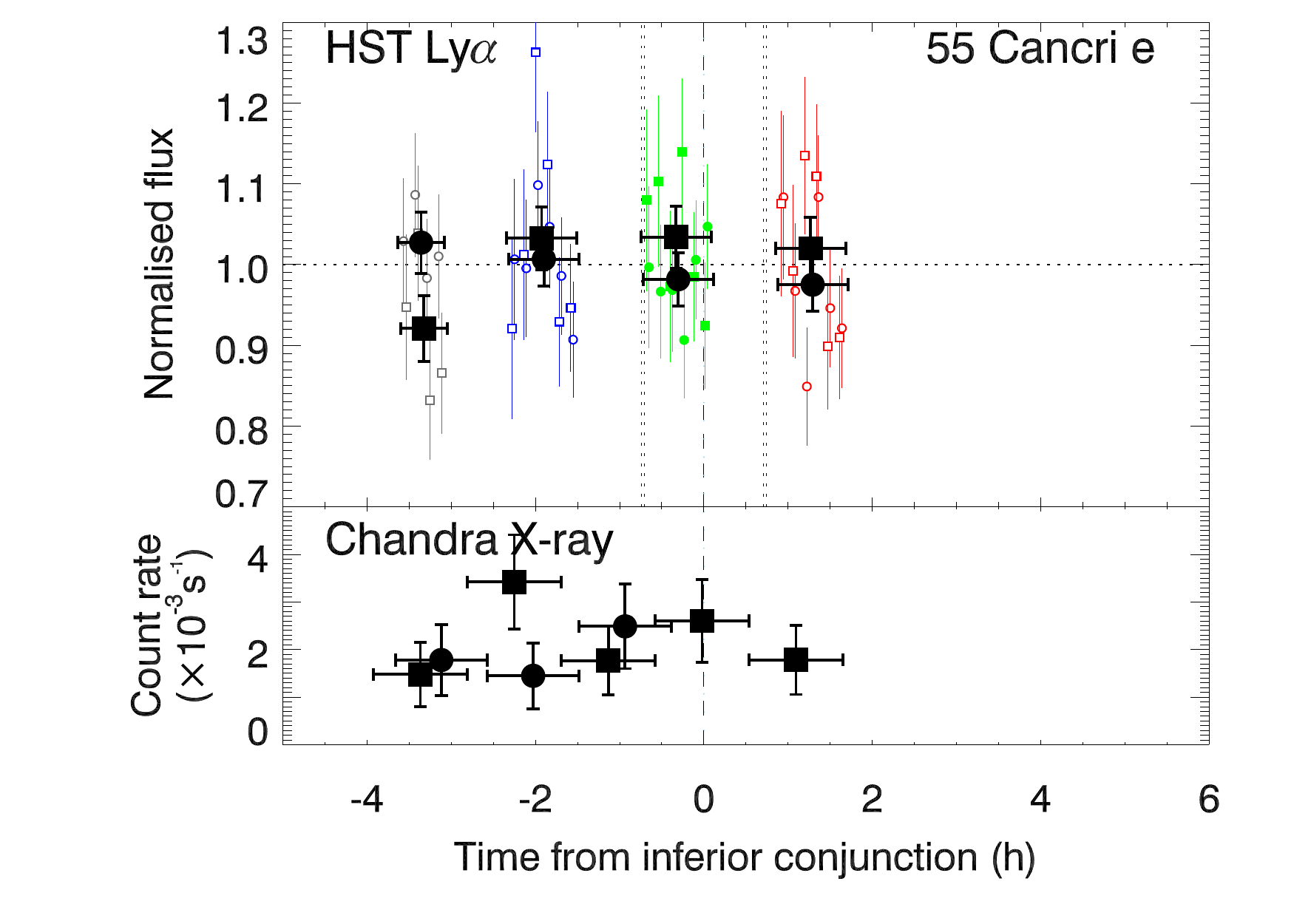}}
\caption{\label{fig:LCe} Same as Fig.~\ref{fig:LCb} for 55~Cnc~e. \emph{Top:} \lya\ light curve integrated between 1\,215.36 and 1\,215.67~\AA. The vertical dashed line indicate mid-transit and the dotted lines the transit contacts. \emph{Bottom:} \emph{Chandra} count rate.}
\end{figure}

As a test, we calculated the transit signal of 55~Cnc~e over the same wavelength interval used for 55~Cnc~b, i.e., between 1\,215.36 and 1\,215.67~\AA. On this interval, where it is possible to measure a transit depth for 55~Cnc~e with the same precision as for 55~Cnc~b, we obtain an absorption depth of $0.3\pm2.4$\%. The corresponding light curve is plotted in Fig.~\ref{fig:LCe} (top panel).

Our conclusion is that we do not detect any transit signal from 55~Cnc~e at \lya.

\subsection{55~Cancri X-ray emission}
\label{sec:chandraresults}
Inspection of the \chandra\ ACIS-S spectra of 55~Cnc shows it to be a very soft source (as expected for a relatively inactive late-type star) with the majority of detected events lying softer than the limit of the \chandra\ ACIS-S spectral calibration (0.243~keV). 
Analysing the spectrum from the longer second observation (using XSPEC version 12.7), we found a marginally acceptable fit using a single-temperature collisionally-ionised plasma model \citep[APEC;][]{Smith:2001}. Given the low number of counts we minimised using the Cash statistic and tested goodness of fit using Monte-Carlo simulations. For a best fitting single temperature of $0.09\pm0.02$~keV we found 80\% of simulated spectra yield lower fit statistics. The $[0.25,2.5]$-keV or $[5,50]$-\AA\ flux was $2.4\times10^{-14}$~erg~s$^{-1}$~cm$^{-2}$, corresponding to an X-ray luminosity in this band of $4\times10^{26}$~erg~s$^{-1}$.

In order to make the best comparison with the X-ray flux at the time of the 
\xmm\ observation in April 2009, we carried out a simultaneous fit of the \xmm\ EPIC-pn spectrum and our second (longer) \chandra\ ACIS-S observation. We used a 
two-temperature APEC model, forcing the temperatures to be the same for the two observations but allowing the normalisations vary indepdendently. It was immediately obvious that the X-ray emission at the time of the \chandra\ observation was both softer and fainter than at the time of the \xmm\ observation. An acceptable fit was found (with 62\% of Monte-Carlo realisations of the model spectrum yielding lower values of the Cash statistic) with best fitting temperatures of $0.09\pm0.01$ and $0.6\pm0.1$~keV. The normalisation of the low-temperature component was a factor 1.7 lower for the \chandra\ data compared to the \xmm\ data, while the normalisation of the high-temperature component was a factor of 45 lower. This suggests that the EUV emission was relatively similar at the times of the two observations, while the X-ray flux was much lower at the time of the \chandra\ observation. Averaged across the $[0.25,2.5]$-keV ($[5,50]$-\AA) band, the X-ray emission was a factor 2.4 lower at the time of our \chandra\ observations than at the time of the \xmm\ observation in April 2009. 

It is clear that 55~Cnc is a variable EUV and X-ray source, with weaker 
irradiation of 55 Cnc e and b in March/April 2012 than in April 2009.
The X-ray flux was consistent between the two \hst\ visits, however,
and no flares were detected, giving us confidence that it is reasonable to 
assume that the mass loss rates from the planets were also similar at the 
two \hst\ epochs. 

Together with the detection of variable mass loss from HD~189733b 
\citep{LecavelierdesEtangs:2012}, our results underline the need for 
future Lyman-$\alpha$ studies of extended planetary atmospheres to be 
accompanied by contemporaneous EUV/X-ray observations. 

\section{Discussion}
\label{sec:discussion}

\subsection{Can 55~Cancri~b not transit whilst hosting a (partially) transiting exosphere?}

We detect a \lya\ absorption signal during inferior conjunctions of 55~Cnc~b, suggestive of a transit-like event. Can it be the signature of a partial or grazing planetary transit with an impact parameter, $b=a_b/R_\star \cos i$, close to 1? If the orbital planes of 55~Cnc planets are all aligned\footnote{\citet{McArthur:2004} tentatively report an inclination of $53\degr\pm6.8\degr$ for the farther out jovian planet \object{55~Cnc~d} from \hst\ astrometry, indicating possible misalignments in the system.}, and considering the inclination of $i_e=82.5^{+1.4}_{-1.3}$~deg derived for 55~Cnc~e by \citet{Gillon:2012} and the semi-major axis of 55~Cnc~b ($a_b=0.11$~\AU\ or $26.19~R_\star$), the latter could have an impact parameter in a range of $[2.8,4.0]$ ($1\sigma$) or $[1.5,5.2]$ (3$\sigma$), therefore excluding a transit for any reasonable values of the planetary radius. Meanwhile, owing to its `large' separation with the star, 55~Cnc~b also has a large Roche radius of $\sim1.2~R_\star$. Consequently, the Roche lobe of 55~Cnc~b \emph{can} undergo partial transits. If it is filled with hydrogen, as observed for hot Jupiters, then a transit signature may be observed at \lya. 

Besides, the value of $i_e$ found by \citet{Gillon:2012} is the least favourable to a grazing transit of 55~Cnc~b, assuming planets b and e are aligned, among the other values found in the litterature \citep{Winn:2011,Demory:2012} or in the recent report by the \emph{MOST} team of an inclination $i_e = 85.4\pm2.5$~deg (D.~Dragomir, 2012-08-27), allowing for an impact parameter of 55~Cnc~b lying between 1.0 and 3.2.

\begin{figure}
\resizebox{\columnwidth}{!}{\includegraphics[trim=2.cm 0.5cm 1cm 0.5cm]{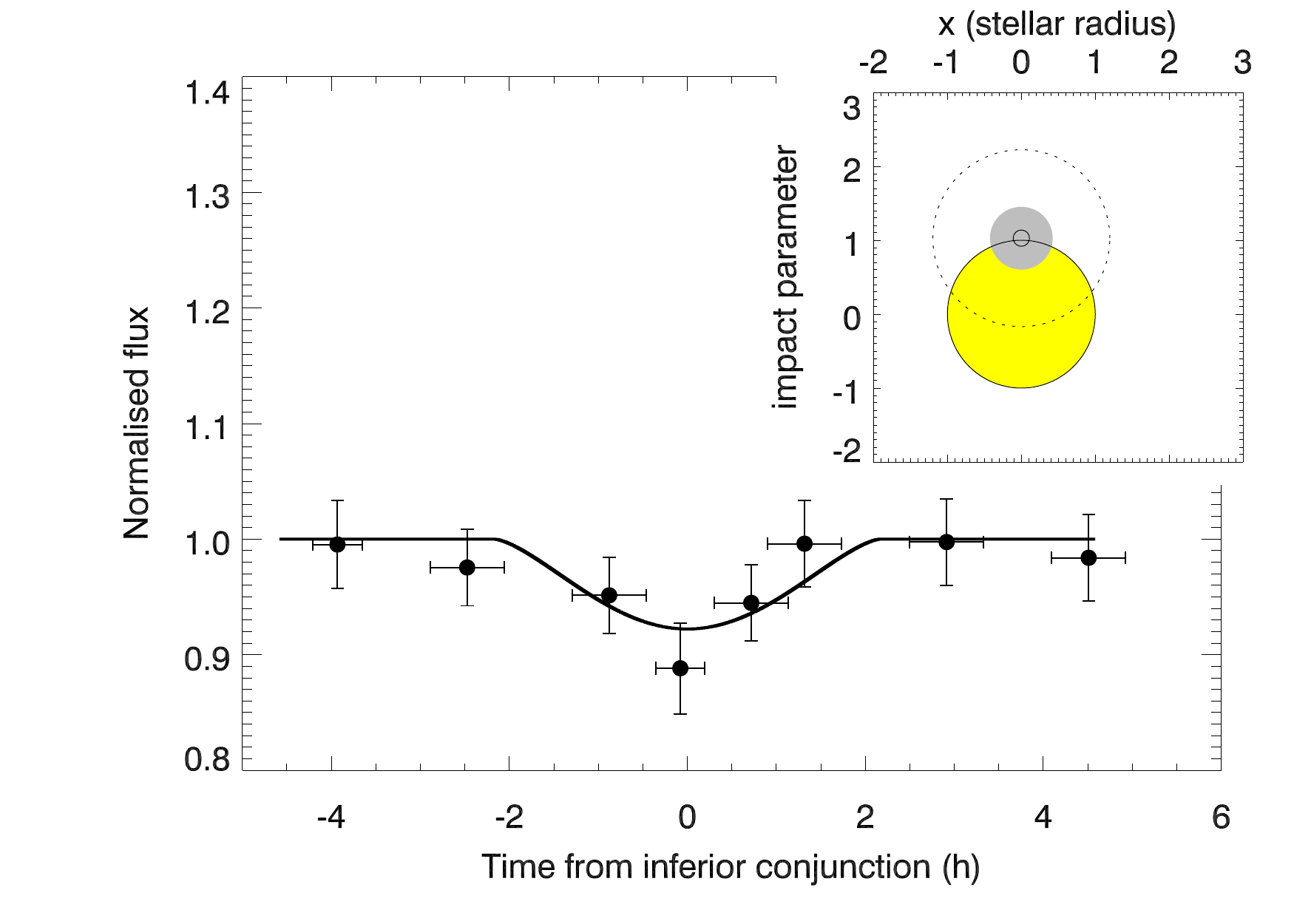}}
\caption{\label{fig:geometry} Lyman-$\alpha$ light curve of 55~Cnc~b from Fig.~\ref{fig:LCb} (top panel) compared to a best-fit model transit light curve (thick line) with an impact parameter of 1.03 and an optically thick radius of 0.35 times the size of the Roche lobe. \emph{Inset:} Sketch of the system at inferior conjunction. The Roche radius of the planet is represented by the dotted circle, while the shaded region corresponds to the optically thick hydrogen envelope.}
\end{figure}

We built a simple model to test this hypothesis. Considering the impact parameter $b$ and the optically thick radius of the hydrogen exosphere $r_\mathrm{exo}$ as free parameters, we fit a transit light curve calculated with the routines of \citet{Mandel:2002} to the data. A best-fit model with a $\chi^2$ of 2.3 for 6 degrees of freedom was obtained for $b=1.03$ and $r_\mathrm{exo} = 0.35$ Roche radius. The result is shown in Fig.~\ref{fig:geometry}.

Assuming 55~Cnc~b has a radius similar to Jupiter, the parameters given above would imply a small partial transit by the planet itself, that would be however challenging to confirmed or ruled out with high-precision photometry. Meanwhile, we emphasise that this problem is underconstrained, since solutions with non-transiting planet b would still provide decent fits to the data. In addition, a few-degree misalignment between the orbital planes of these two planets cannot be ruled out. Confidently excluding grazing transits of 55~Cnc~b with precise photometric observations could bring a useful insight on this question.

The simple model described above does not take into account the dynamics of hydrogen atoms. Looking at the transmission spectrum from Fig.~\ref{fig:Spectra_b}, the absorption signal takes place in the blue part of the \lya\ line, between 0 and $-76.5$~\kms, as expected in the case of neutral hydrogen atoms repelled by the stellar radiation pressure towards the observer. In order to assess whether radiation pressure can reproduce the observed velocity distribution of the absorption signal, we calculated the 2-D trajectory of a hydrogen atom released from the upper atmosphere of 55~Cnc~b with a tangential speed equal to the orbital velocity of the planet. The particle is subject to the stellar gravity and radiation pressure. The latter is estimated by reconstructing the intrinsic stellar \lya\ line from the observed profile as shown in Fig.~\ref{fig:Spectra_b}, following the method of \citet{Ehrenreich:2011a}. 

\begin{figure}
\resizebox{\columnwidth}{!}{\includegraphics[trim=0.5cm 0.5cm 1cm 0.5cm]{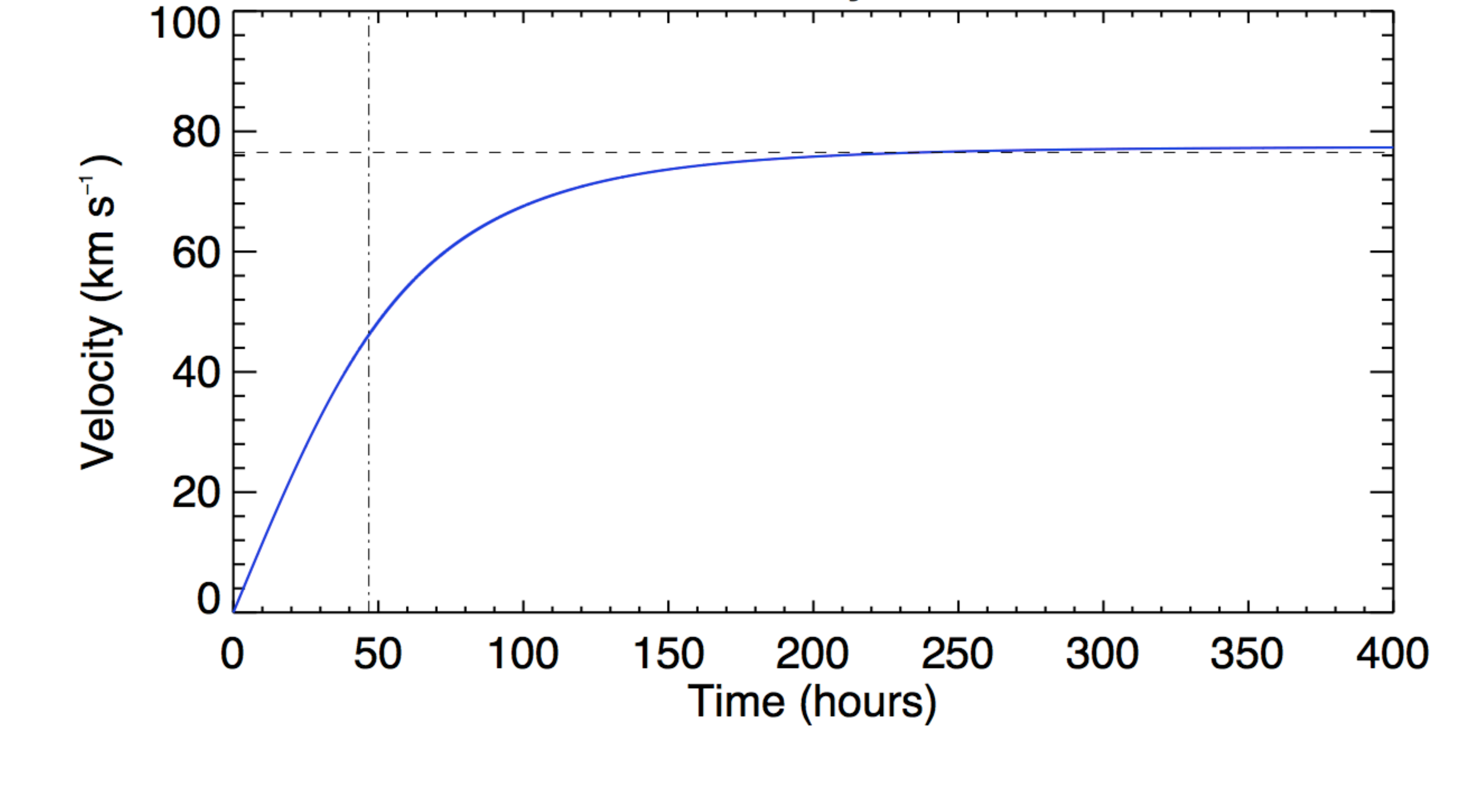}}
\caption{\label{fig:rvtime} Plot of the radial velocity of a hydrogen atom launched from 55~Cnc~b in a simplified 2D simulation (solid blue line). The horizontal black dashed line shows the observed absorption limit velocity of $\sim76.5$~\kms. The vertical blue dashed line shows the characteristic life time of a hydrogen atom for an ionising flux of 1 solar EUV flux.}
\end{figure}

The ratio $\beta$ between radiation pressure and gravity \citep{Lagrange:1998} is reported as a function of radial velocity on the right axis of Fig.~\ref{fig:Spectra_b} (top panel). Its value close to 1 between $-80$ and $+80$~\kms\ and its steep decrease at larger velocities (combined with the planet's orbital velocity of 85~\kms) implies that as soon as the atoms escape the planet, they are rapidly accelerated by radiation pressure at negative radial velocity down to $-76.5$~\kms\ in $\sim120$~h, as shown in Fig.~\ref{fig:rvtime}. This acceleration time scale is just 2 to 3 times larger than the characteristic time scale of ionisation of neutral hydrogen by EUV stellar photons, which is $\tau\sim50$~h assuming a solar EUV flux, and we predict that escaping atoms must present a distribution of radial velocities ranging from $0$ to $-76.5$~\kms, in agreement with the limit of the detected absorption.\footnote{After a time $t$, there will be a fraction of $\exp(-t/\tau)$ of neutral atoms remaining. After 120~h, there will thus be $\sim10$\% of neutral atoms with a radial velocity of $-76.5$~\kms.} Finally, we also note that the absence of any signal towards the red is consistent with the radiation pressure interpretation.

We stress that additional data will be mandatory to confirm this result. Since the star is variable in X-rays (Sect.~\ref{sec:chandraresults}), similar observations at a different epoch could result in an increased transit signal, or no signal, as seen for HD~189733b \citep{LecavelierdesEtangs:2012}, pleading for new simultaneous \lya\ and X observations. If the transit signal of 55~Cnc~b is confirmed, subsequent modelling efforts should help understanding the depth and duration of the transit light curve assuming atmospheric escape rate much lower than what is currently known for hot Jupiters (see Sect.~\ref{sec:discuss-e}), taking into account the roles of radiation pressure, ionisation, self-shielding, as well as stellar wind interactions as another possible acceleration mechanism of neutral H atoms at large distances from the planet \citep{Holmstrom:2008a,Ekenback:2010,Tremblin:2012}.

In particular, the light curve tentatively hints at an early ingress, not unlike what was claimed to be seen during the UV transit of the hot Jupiter Wasp-12b \citep{Fossati:2010,Vidotto:2010}.

\subsection{Implications of the non-detection of 55~Cnc~e at Lyman~$\alpha$}
\label{sec:discuss-e}

The absence of a \lya\ detection at the transit times of 55~Cnc~e implies that if this planet has retained a thick atmosphere, its spatial extent has to be limited. Given its reported radius and mass, 55~Cnc~e has a mean density of $\rho_e = 4.0^{+0.5}_{-0.3}$~g~cm$^{-3}$ \citep{Gillon:2012}. This is too low for this planet to be telluric, i.e., a scaled-up version of the Earth mainly consisting of an iron core and a rocky mantle. Instead, internal structure models presented in \citet{Gillon:2012} suggest that a large fraction ($\sim20\%$) of volatile is needed to account for the mass and size of the planet. One possibility is that the planet is enshrouded in a thick envelope of supercritical ($> 647$~K) steam. This water envelope could then provide a large source of atomic hydrogen under the photodissociated effect of the incoming stellar EUV and X-ray radiation. In this scenario, inspired by the prediction of the observational signature of `evolved oceans' on terrestrial exoplanets \citep{Jura:2004}, a \lya\ transit signature could be expected depending on the atmospheric mass loss rate and the stellar photodissociating flux.

We scaled the atmospheric escape model used in previous studies of hot Jupiters or Neptunes \citep{Vidal-Madjar:2003,Ehrenreich:2011a,LecavelierdesEtangs:2012} to the orbital and physical properties of 55~Cnc~e (a full description of the model will be provided in Bourrier et al., in preparation). Our non-detection allows us to set a 3-$\sigma$ upper limit of $\la3\times10^8$~g~s$^{-1}$ to the atmospheric mass loss rate of 55~Cnc~e.

Taking the stellar X-ray luminosity from Sect.~\ref{sec:chandraresults} we can calculate X-ray fluxes at 55~Cnc~e and b respectively of 640 and 12~erg~s$^{-1}$cm$^{-2}$ (in the 0.25--2.5~keV band). These fluxes correspond to energy-limited escape rates of $\sim5\times10^8$ and $2\times10^7$~g~s$^{-1}$, respectively, using the prescription of \citet{LecavelierdesEtangs:2007} and \citet{Ehrenreich:2011b}. We note here that the energy-limited escape rate estimated for 55~Cnc~b is much weaker than the rates of $\sim$$10^9$ to $10^{10}$~g~s$^{-1}$ observationally derived for the hot Jupiters HD~209458b or HD~189733b.

For 55~Cnc~e, our upper limit to mass loss is lower than the X-ray energy-limited rate, even without considering the EUV flux which may be an order of magnitude higher \citep[e.g.][]{Sanz-Forcada:2011}. This appears to rule out high mass loss efficiencies ($\ga 60\%$).

A mass loss rate of $3\times10^8$~g~s$^{-1}$ is insufficient to completely erode a putative water envelope representing 20\% of 55~Cnc~e in mass. A rough yet conservative calculation, unphysically assuming that all the water (i.e., more than 1~\Mearth) has been photodissociated into hydrogen, yields a value of $\ga 6\times10^{10}$~g~s$^{-1}$ for a complete atmospheric escape to occur within 5~Gyr. This is two order of magnitude larger than our 3-$\sigma$ upper limit, indicating that a massive photodissociating water envelope scenario is not ruled out by our observations. At this stage, however, we note that a proper calculation would require a photochemical model for the upper atmosphere of a super-Earth such as 55~Cnc~e.

Finally, another interesting but more exotic scenario could be that of a thick carbon dioxide atmosphere on 55~Cnc~e instead of a water envelope. In this case, we would not expect to observe a \lya\ signature during transits.

\section{Conclusion}

Our \hst\ and \chandra\ observations reveal that 55~Cnc is bright at \lya\ and variable in X-rays. This opens interesting perspectives for testing the impact of the incident X-ray and UV radiation on the atmospheric heating and chemistry for planets in this system, which cover broad ranges of both mass and separation \citep[see, e.g., the study of the \object{GJ~876} system by][]{France:2012}.

No \lya\ transit signals are detected for 55~Cnc~e, allowing us to estimate an upper limit to the escape rate of a gaseous envelope enshrouding this super-Earth. This limit, of $\sim$$3\times10^8$~g~s$^{-1}$, does not preclude the existence of such a volatile envelope.

We found that the \lya\ stellar emission is absorbed at times corresponding to the inferior conjunction of 55~Cnc~b. Furthermore, the absorption is blueshifted and observed down to $\sim-76.5$~\kms\ from the line centre, as expected from atoms expelled by the stellar radiation pressure. Since the planetary system is close to being seen edge-on, this hints at the tantalizing possibility of a partially transiting extended \ion{H}{i} atmosphere around 55~Cnc~b, escaping the planet at a rate of 2 to 3 orders of magnitudes lower than what is known for hot Jupiters.

Yet, additional observations at \lya, as well as in other lines (e.g., \ion{C}{ii}, \ion{Mg}{ii}, or \ion{Na}{i}) are clearly needed to confirm this result. Meanwhile, it pushes to continue the photometric follow-up of 55~Cnc, to firmly rule out grazing transits of 55~Cnc~b in the visible, in case the orbital planes of planets b and e are slightly misaligned.

If a transiting extended atmosphere on 55~Cnc~b is confirmed, it would represent the first atmospheric characterisation of a giant planet that is not a hot Jupiter: as such, it would contribute to bridge the gap between hot exoplanets and the cooler planets in our Solar System, which experienced hydrodynamical atmospheric escape early in their history. The bright star 55~Cnc would then become our best laboratory for these comparative studies. 

\acknowledgements
We thank STScI and \chandra\ directors M.~Mountain and H.~Tananbaum, for the allocation of DD time on \hst\ and \chandra, as well as our \hst\ programme coordinator T.~Royle, and contact scientist R.~Osten for tremendous support. We are also grateful to D.~Dragomir for helpful discussions about planet~b and the referee I.~Snellen for thoughtful comments that helped improve the manuscript. This work was made possible with the funding of the Grand Prix Grivet of the Acad\'emie des sciences and support from CNES. DE also acknowledges the funding from the European Commission's Seventh Framework Programme as a Marie Curie Intra-European Fellow (PIEF-GA-2011-298916). This work is based on observations made with the NASA/ESA \emph{Hubble Space Telescope}, obtained at the Space Telescope Science Institute, which is operated by the Association of Universities for Research in Astronomy, Inc., under NASA contract NAS 5-26555.

\bibliographystyle{aa}
\bibliography{biblio.bib}

\end{document}